\documentclass[twocolumn,onecolappendix,tighten,times]{aastex62}

\usepackage{tikz}
\usepackage{amsmath}
\usepackage{multirow}
\usepackage{nicefrac}
\usepackage{enumitem}
\usepackage{pbox}
\usepackage{soul}
\usepackage{float}
\usepackage{lineno}

\let\OLDthebibliography\thebibliography
\renewcommand\thebibliography[1]{
  \OLDthebibliography{#1}
  \setlength{\parskip}{-1pt}
  \setlength{\itemsep}{-1pt plus 0.3ex}
}

\shorttitle{Exoplanet Radius Cliff Explained by Fugacity Crisis}
\shortauthors{Kite et al.}

\begin{document}


\title{\Large{\textbf{Superabundance of Exoplanet Sub-Neptunes Explained by Fugacity Crisis}}}%


\author{Edwin S. Kite}%\altaffilmark{1}}
\affil{Department of the Geophysical Sciences, University of Chicago, Chicago, IL (kite@uchicago.edu).}
%\email{kite@uchicago.edu}

\author{Bruce Fegley Jr.}%\altaffilmark{4,5}}
\affil{Planetary Chemistry Laboratory, McDonnell Center for the Space Sciences \& Department of Earth \& Planetary Sciences, \\ Washington University, St Louis, MO .}

\author{Laura Schaefer}%\altaffilmark{5}}
\affil{School of Earth Sciences, Stanford University, Palo Alto, CA.}
\author{Eric B. Ford}\affiliation{Department of Astronomy and Astrophysics, The Pennsylvania State University, University Park, PA }
\affiliation{Center for Exoplanets and Habitable Worlds, The Pennsylvania State University, University Park, PA}
\affiliation{Institute for CyberScience}
\affiliation{Pennsylvania State Astrobiology Research Center}

%Center for Exoplanets \& Habitable Worlds, Department of Astronomy \& Astrophysics; \\
%Center for Astrostatistics; \& Institute of CyberScience,

%\email{kite@uchicago.edu}

\begin{abstract}
\noindent Transiting planets with radii 2-3 $R_\earth$ are much more numerous than larger planets. We~propose that this drop-off is so abrupt because at $R$~$\sim$~3~$R_\earth$, base-of-atmosphere pressure is high enough for the atmosphere to readily dissolve into magma, and this sequestration acts as a strong brake on further growth. The viability of this idea is demonstrated using a simple model. Our results support extensive magma-atmosphere equilibration on sub-Neptunes, with numerous implications for sub-Neptune formation and atmospheric chemistry. 
 \end{abstract}
\vspace{0.1in}


\section{Introduction.}
\noindent

\begin{figure}
\centering
\includegraphics[width=1.0\columnwidth,clip=true,trim={10 0 55 40}]{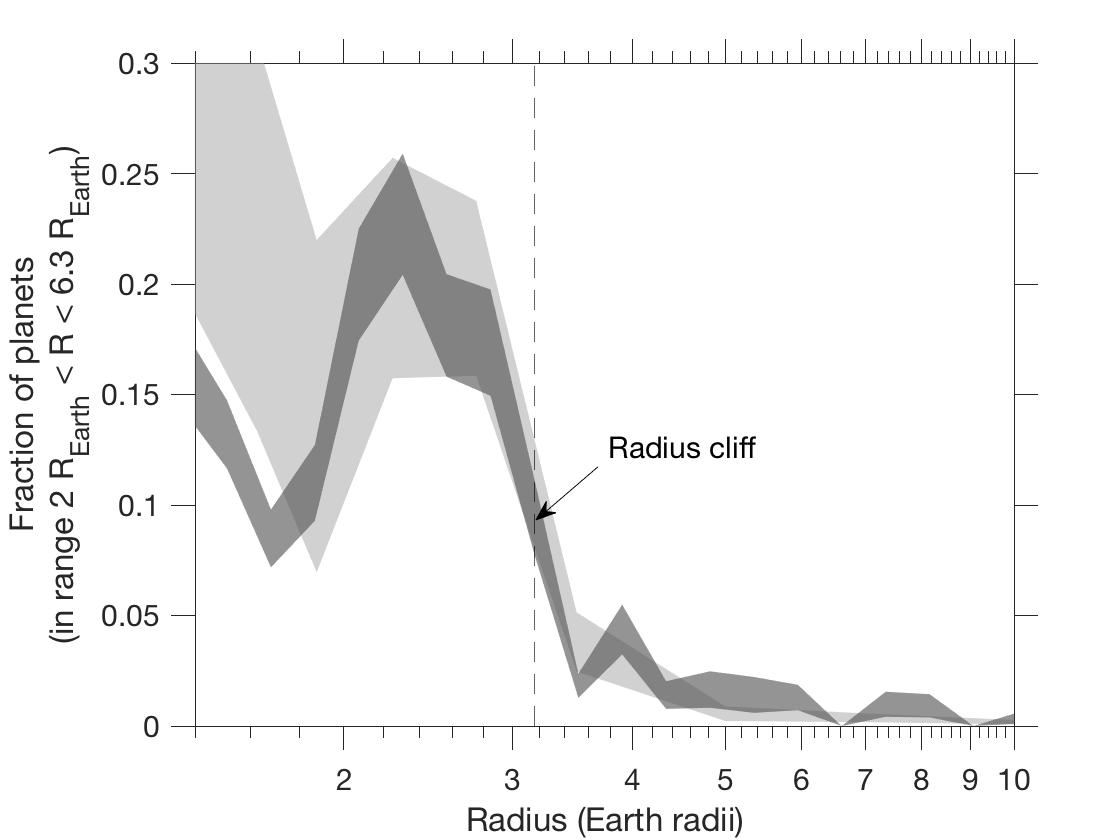}
\caption{The exoplanet radius distribution, according to \citet{FultonPetigura2018} (dark gray band, $\pm$1$\sigma$), and according to \citet{Hsu2019}, (for $p$~$<$~64~days, light gray band, $\pm$1$\sigma$). \citet{Hsu2019} data are adjusted downward by a factor of 2.25 in order to compensate for different bin choices. The dashed line at 3.16~$R_\Earth$ highlights the radius cliff.}  
\label{fig:keplercliff}
\end{figure}

\noindent According to \emph{Kepler} data, the intrinsic frequency of 2.7-3.0~$R_\earth$ planets is (4-10)$\times$ that of planets that are only 20\% bigger (3.3-3.7~$R_\earth$) (for $p$~$<$~100~d; \citealt{FultonPetigura2018,Hsu2019}). This drop-off, or cliff, is the most dramatic feature in the planet-radius histogram (Fig.~\ref{fig:keplercliff}). The~radius~cliff separates sub-Neptunes, which are intrinsically common, from intrinsically rare Neptune-sized and larger planets. What censors planet growth beyond $\sim$3 $R_\earth$? Here we~propose a new explanation for the steepness (amplitude) and location of the radius cliff. We~attribute both to increased solubility of hydrogen in magma at pressures $>$10$^9$~Pa. We~assign this consequence of non-linear hydrogen solubility for sub-Neptune radii a short-hand name, the \emph{fugacity crisis}.

The \emph{Kepler} sub-Neptunes are made of Earth-composition cores (silicates, plus Fe-metal) shrouded by hydrogen-dominated atmospheres, based on strong (though indirect) arguments (e.g., \citealt{Rogers2011,OwenWu2017,vanEylen2018,JinMordasini2018,Carrera2018}). We~will accept (for the purposes of this paper) those arguments, which imply that \emph{Kepler} sub-Neptunes are mostly core by~mass, and mostly atmosphere by volume. Because the cores only rarely exceed $\sim$20\% of sub-Neptune volume, the~cliff must correspond to a cut-off in atmosphere volumes. Atmosphere volume is a proxy for atmosphere mass $M_{atm}$\citep{LopezFortney2014}, so the cliff signifies an upper limit on $M_{atm}$ of \emph{O}(2~wt\%) of core mass ($M_{core}$). This is much less than the \emph{O}(100~wt\%) associated with runaway growth into a gas giant \citep{Pollack1996}. As a~result, the upper limit on atmosphere masses cannot be simply explained by core accretion runaway.

Previous attempts to explain the radius cliff have considered both H$_2$ accretion and H$_2$ loss. For a given planet mass, large atmospheres are more weakly bound, and lost more readily (e.g.~\citealt{OwenWu2017}). However, it is unclear if the steepness of the cliff can be explained by H$_2$ loss given the wide range of measured sub-Neptune masses  (e.g., \citealt{Rice2019}). In an alternative model by \citet{LeeChiang2016}, the role of atmosphere accretion is emphasized. In this model, cores -- which are treated as chemically and thermally inert -- receive an atmosphere transfusion from protoplanetary disks just as those disks expire; because gas is sparse during this brief epoch, the atmosphere dose is \emph{O}(2~wt\%). This model may help to explain the scarcity of $p$~$<$~100~d gas giants and matches pre-2018 data. However it is dependent on disk/nebula-era transients, and because disks vary in their properties and lifetimes, it is difficult to see how it can be solely responsible for the steepness of the cliff (Fig.~\ref{fig:keplercliff}). Moreover, the assumption of chemically and thermally inert magma is questionable (e.g., \citealt{Ginzburg2018,Vazan2018}). Hence we seek an alternative explanation.

We are prompted to seek such an explanation in the material properties of H$_2$, specifically the solubility of H$_2$ in magma. The pressure at the atmosphere-core boundary on sub-Neptunes is $P_{atm} \approx M_{atm} \overline{g} / A_{pl}$, where $\overline{g}$~is the magnitude of gravitational acceleration in the atmosphere (using a mass-weighted average), and $M_{atm} \ll M_{core}$. So, if $\overline{g} = \epsilon GM_{core}/R_{core}^2$ (where $R_{core}$ is core radius), then $P_{atm} \approx \epsilon M_{atm} (G M_\Earth / 4 \pi R_\Earth^4 )$, where $\epsilon < 1$ is a correction for lower gravity higher in the atmosphere. Here we set $R_{core}/R_\Earth \sim (M_{core}/M_\Earth)^{1/4}$ (cores are modestly compressible; \citealt{Valencia2006}). This yields

\begin{equation}
P_{atm} \approx \mathrm{5\times10^9 \,Pa} \,\epsilon \left( \frac{f_{atm}}{ 0.01} \right)  \left( \frac{M_{core}}{4 M_\Earth} \right)  
\end{equation}

\noindent  where $f_{atm} = M_{atm} / M_{core}$. Such deep atmospheres slow the cooling of initially molten planetary cores, so most transiting sub-Neptunes will still have a magma ocean in contact with the atmosphere, defining a \emph{magma-atmosphere interface} at which solubility equilibrium should hold. For H$_2$ solubility, 5~GPa is an interesting number. Above 1~GPa, intermolecular repulsion renders molecular H$_2$ much less compressible \citep{Saumon1995}. (Non-ideal behavior kicks in at much lower pressure than the transition to metallic hydrogen, which occurs at $\gtrsim$100~GPa within planets). The reduced compressibility of molecular H$_2$ greatly increases the tendency of H$_2$ to dissolve into adjacent liquid -- this  tendency is termed fugacity ($f$, units~Pa). To understand this, consider the equation for Gibbs free energy $G$, d$G$ = $P$d$V$ - $T$d$S$. If~we assume isothermal conditions and if $\Delta V$ = $V_{\mathrm{(H2,\, in\, melt)}}$ - $V_{\mathrm{(H2, gas)}}$ is negative, then $\Delta G$ favors dissolution. For example, at (1.5-3)~GPa, the density of H$_2$ gas is 40-80~kg/m$^3$, much less than the 180~kg/m$^3$ partial density of H$_2$ in basaltic melt \citep{Hirschmann2012}. As long as the dissolved H$_2$ compressibility exceeds that of the gas at higher $P$, dissolution remains favored. Even for the $f = p$ limit, which is appropriate for $<$1~GPa, the H content of the magma can exceed the H content of the atmosphere \citep{ChachanStevenson2018}. Above 1~GPa, $f_{\mathrm{H2}}$~$\gg$~$P_{\mathrm{H2}}$ (Fig.~\ref{fig:fugacitycrisis}a). This suggests that the ramp-up in dissolution of the atmosphere into magma for  $P_{atm}$~$>$~1~GPa might lead to greater and greater partitioning of added nebula gas into the magma as planet radius (and thus atmospheric mass) increases. We term this a \emph{fugacity crisis} (Fig.~\ref{fig:fugacitycrisis}b). 

\section{Method.}

We seek to explain the 3~$R_\Earth$ cliff, not the divot (``radius valley'') at $\sim$2 $R_\Earth$. Previous studies have proposed explanations for the radius valley, including gas escape-to-space (e.g.~\citealt{OwenWu2017,Ginzburg2018,GuptaSchlichting2019}). We neglect gas escape-to-space, and our model is not intended to match the radius valley.

\begin{figure}

\centering

\includegraphics[width=1.0\columnwidth,trim={0mm 0mm 0mm 0mm}]{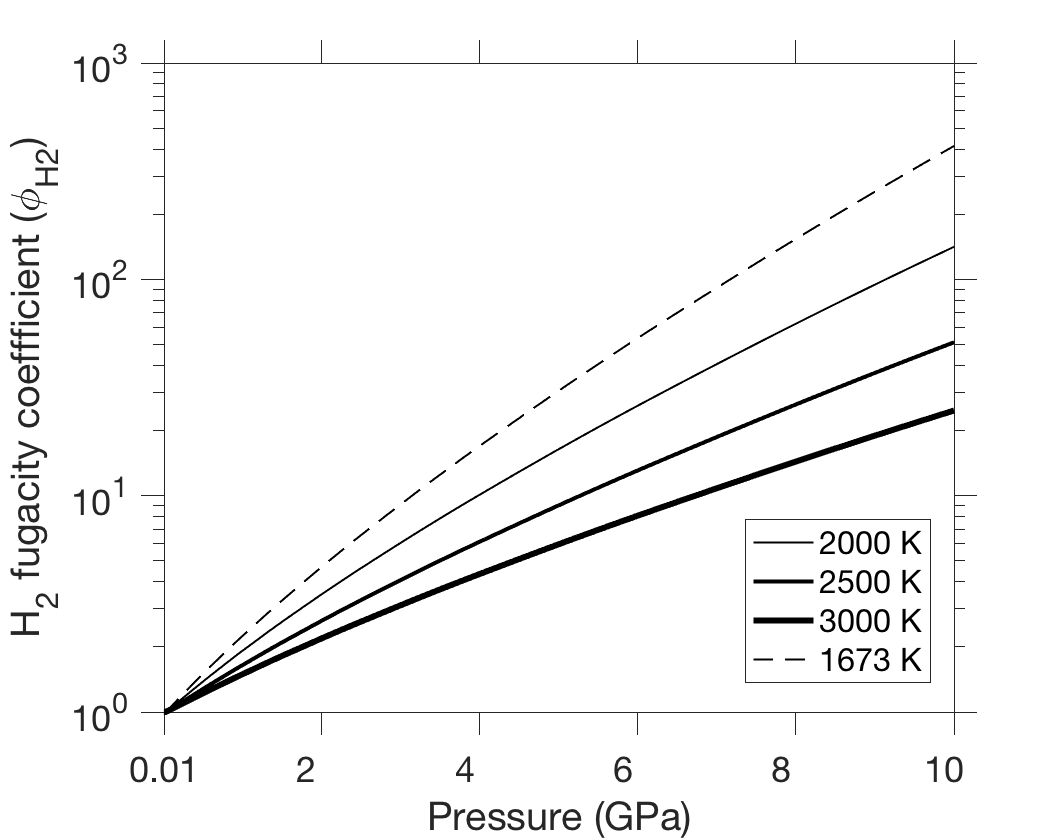}
\includegraphics[trim={0mm 0mm 20mm 10mm},clip,width=1.05\columnwidth]{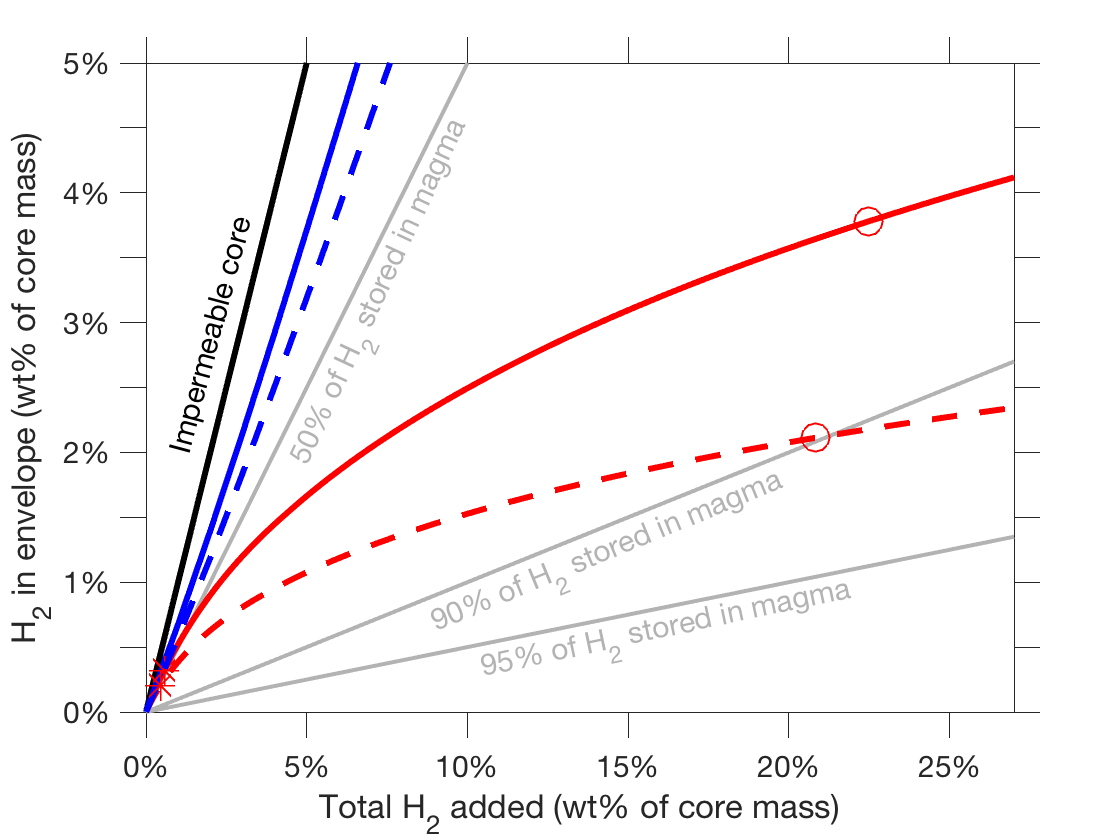}
\caption{(a) Fugacity coefficients for H$_2$. (b) The fugacity crisis for magma-atmosphere interface temperature ($T_{mai}$)~=~3000K. The 1:1 line is the inert impermeable core assumption (``all-H$_2$-in-atmosphere''), used in all but one previous study. The thick blue lines are for the ideal (Henry's Law) dissolution case. The red lines include nonideal dissolution of H$_2$ into magma. Red asterisks show $P_{atm}$~=~1~GPa and red open circles show $P_{atm}$~=~8~GPa. The solid lines show $M_{core}$~=~4~$M_\Earth$. The dashed lines show $M_{core}$~=~8~$M_\Earth$.}  
\label{fig:fugacitycrisis}
\end{figure}

To see if the fugacity crisis can generate a radius cliff, our minimal model includes: 

(a) H$_2$ solubility as a function of $P_{atm}$ and the temperature at the magma-atmosphere interface ($T_{mai}$) (Fig.~\ref{fig:fugacitycrisis}a);

(b) $P_{atm}$ as a function of atmosphere mass;

(c) An expression for the mass of magma as a function of planet mass, tracking magma crystallisation at low $T_{mai}$.

\vspace{0.05in}

Details of (a)-(c) are provided below.  By combining (a)-(c), we can calculate the partitioning of H$_2$ between the magma and the atmosphere (Fig.~\ref{fig:fugacitycrisis}b).

\vspace{0.05in}

(a) Getting H solubility as a function of pressure and temperature requires fugacity coefficients, an experimental fugacity$\leftrightarrow$solubility calibration, and a temperature dependence parameterization \citep{Kite2019}. The fugacity coefficient of H$_2$, $\phi$, is computed using

\begin{equation}
\mathrm{ln}\phi= \mathrm{ln} \frac{f}{P}= \int_0^P \left(\frac{Z-1}{P}\right) dP
\label{eqn:fugacityfromz}
\end{equation}

\noindent where the compressibility factor $Z$ is given by $Z = (PV_m)/(RT)$. Here, $V_m$ is the molar volume, obtained from the \citet{Saumon1995} tables assuming pure molecular H$_2$ (we set $Z$~=~1 below 10$^7$~Pa to minimize the effect of thermal dissociation). When $\phi$~=~1, the gas behaves ideally. $\phi$ $>$10  by 8~GPa (Fig.~\ref{fig:fugacitycrisis}a). The H$_2$ solubility at the magma-atmosphere interface is set to 

\begin{equation}
X_{\mathrm{H2}}~=~1~\times~10^{-11}\, f_{\mathrm{H2}} \, \mathrm{exp}(-T_{\mathrm{0}}/T_{mai})  
\end{equation}

\noindent where $X_{\mathrm{H2}}$ is the mass fraction. ($X_{\mathrm{H2}}$ is not permitted to exceed 50~wt\%.) This follows the estimated molten-average-rock solubility from \cite{Hirschmann2012} (i.e., the estimated peridotite solubility). This solubility is $\sim$5$\times$ lower than was used by \cite{ChachanStevenson2018}. $T_{\mathrm{0}}$ is uncertain; we use 4000~K (following \citealt{ChachanStevenson2018}). There are no direct measurements of H$_2$~solubility in magma at $\sim$3000~K.

(b) To relate $f_{atm}$ to $P_{atm}$, for $P_{atm}$ $<$ 100~bars we assume $\overline{g} = GM_{core}/R_{core}^2$. For thicker atmospheres, we use the hydrogen equation-of-state of \citet{Saumon1995} to construct adiabatic density-height ($\rho$-$z$) profiles up from the bottom-of-atmosphere temperature (assumed equal to $T_{mai}$) in order to obtain $\overline{g}$. We multiply densities by 120\% to account for non-H$_2$ species. This approach is intended only to make a~first-order correction for the reduced gravity within the atmosphere, and we do not use the output to get the planet transit radius. We also ignore the top-of-atmosphere $T$ from this workflow (typically 500-1500~K at 100 bars). Instead, we treat $T_{mai}$ as a free parameter (sub-Neptune core cooling calculations output $T_{mai}$~=~3000$\pm$1500~K; e.g., \citealt{HoweBurrows2015,Vazan2018, Bodenheimer2018}). 

(c) Molten rock can store very much more volatiles than solid rock. To obtain the mass of rock that is molten and so can store H$_2$, we follow \citet{Kite2019}. Within a convecting magma ocean, $\partial T_{melt} / \partial z$ $>$ $\partial T_{adiabat} / \partial z$ for $P$~$<$~10$^2$~GPa. Here, $T_{melt}$ is the temperature corresponding to 40 wt\% melt fraction, and $T_{adiabat}$ is the temperature within the convecting magma. Thus, sub-Neptunes plausibly have magma shells overlying solid silicates. To find magma shell thickness, we interpolate and extrapolate the solidus (0\% melting curve), the liquidus (100\% melting curve), and the magma adiabats of Figure~5 from \citet{Andrault2011}. We integrate down from the magma-atmosphere interface until the adiabat reaches the solidus. To do this, we extrapolate the silicate density as a function of pressure from \citet{DziewonskiAnderson1981}. The $T_{mai}$~=~3000~K adiabat is hotter than the solidus for chondritic-primitive-mantle material for $P$~$<$~130~GPa according to \citet{Andrault2011}. We make the approximation that at $T_{mai}$ $\ge$ 3000~K, sub-Neptune silicates are fully molten. We assume that silicates make up 2/3 of the mass of the planet core. This very basic model is sufficient for our purposes; see (e.g.) \citet{Bower2019} and \citet{Dorn2017} for more sophisticated models.

We compute planet radii based on $f_{atm}$, using the planet transit radius look-up tables of \citet{LopezFortney2014}, for solar-composition opacity and planet age $\ge$1~Gyr. 

So far, we described calculations for one planet, but our goal is to compare to the planet radius histogram. In order to generate synthetic planet histograms, we need a prior distribution on the variability of the total amount of H$_2$ supplied by the nebula to the core. As shown in Fig.~\ref{fig:sensitivity}, the existence and approximate location of the cliff has low sensitivity to reasonable variations for the choice of prior.

\section{Results.}

\subsection{The Crisis in H$_2$ Partitioning}
The crisis in H$_2$ partitioning is shown in Fig.~\ref{fig:fugacitycrisis}b. For $<$0.5~wt\% of H$_2$ added, for $T_{mai}$ = 3000~K and a 5~$M_\Earth$ core, most of the H$_2$ stays in the atmosphere. However, as the total H$_2$ added is increased, it becomes very difficult to~increase the mass of H$_2$ in the atmosphere because solubility increases exponentially with $P_{atm}$ (Fig.~\ref{fig:fugacitycrisis}a). This is the fugacity crisis. For 10~$M_\Earth$ cores, exceeding 1.5 wt\% H$_2$ in the atmosphere requires $>$20\% H$_2$ to be added, and beyond this point almost all of each additional parcel of H added goes into the core.

\subsection{The Fugacity Crisis Can Explain the Radius Cliff At 3~$R_\earth$}

\begin{figure}
\includegraphics[width=1.05\columnwidth,clip=true,trim={10 0 55 40}]{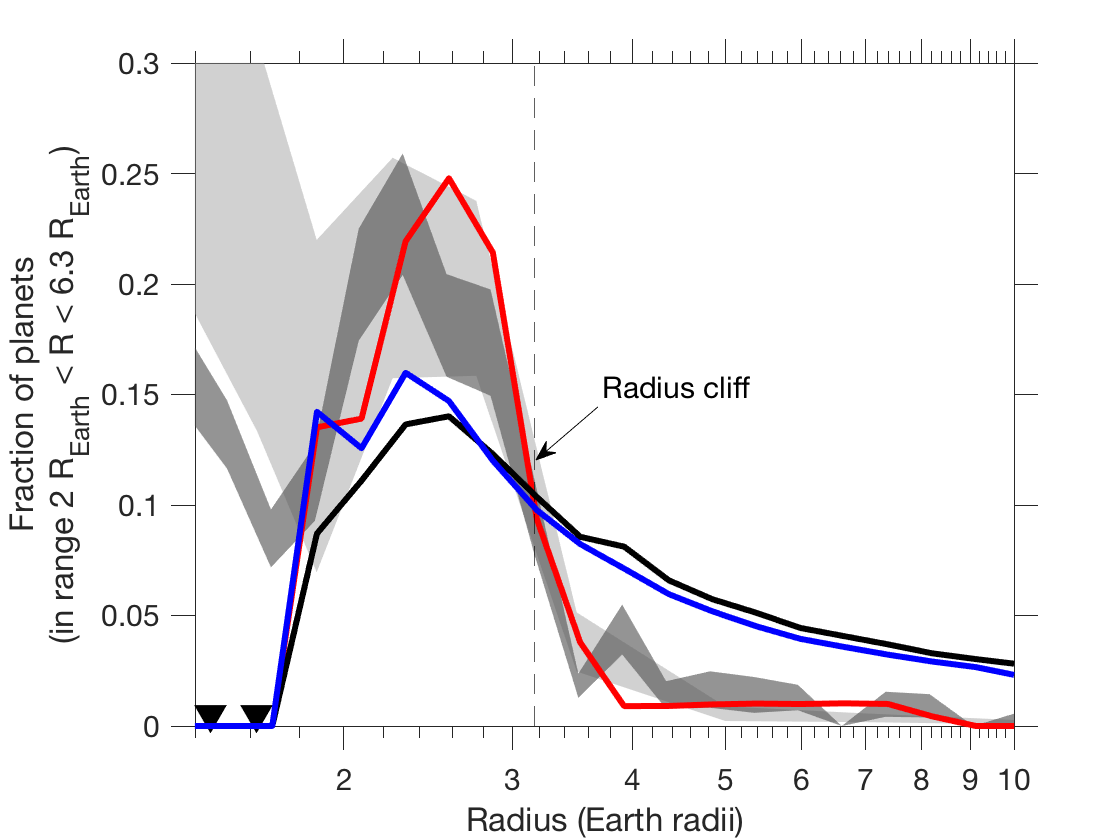}
\caption{Histograms of planet abundance. Colored bands are the true planet histogram ($\pm$1$\sigma$ error) according to  \citet{Hsu2019} (light gray) and according to \citet{FultonPetigura2018} (dark gray). Lines show model output for the impermeable planet case (black line); linear (Henry's Law) dissolution (blue line); and the fugacity-crisis case (red line). Parameters: $M_{core}$ drawn with equal likelihood from \{4,5,6\} $M_\Earth$, $T_{mai}$~=~3000~K, solar-composition atmospheric opacities, insolation 1000 $L_{\Earth}$, planet age 1 Gyr, and a log-gaussian distribution of gas supply centered on 5~wt\% of core mass, with a standard deviation of 1 dex in gas supply, and an upper limit of 50~wt\% (above this limit, we implicitly assume, gravitational runaway will cause planets to explode into exo-Jupiters.) Model output bins are the same as in \citet{Fulton2017}.Triangles correspond to the bare-core radii for 4 and 6~$M_\Earth$. }  
\label{fig:histogram}
\end{figure}

Fig.~\ref{fig:histogram} shows the reference results. With a smooth distribution of gas supply, both the impermeable-core case (black line) and the linear-solubility, Henry's Law case (blue line) yield a broad distribution for radii. Neither model predicts a cliff. 

However, the observed radius cliff is reproduced by the fugacity crisis model. Below $\sim$2.2 $R_\Earth$ (corresponding to  1~GPa), non-ideal effects are small, and the red line closely tracks the blue line. Between 1~GPa and 8~GPa (radius 2.2-3.6 $R_\Earth$) the non-ideal effects are so strong as to define a sharp concentration and a sharp fall-off in planet radii. Essentially, transiting planets with radii 2-3~$R_\Earth$ are so numerous because at
$R$~$\sim$~3~$R_\Earth$, base-of-atmosphere pressure becomes large enough for
the atmosphere to readily dissolve into magma. This sequestration greatly slows the rate of growth in planet radius, even as the planet continues to accrete gas.

All three models shown in Fig.~\ref{fig:histogram} underpredict the inferred planet occurrence rate for planets smaller than $\sim$1.8~$R_\Earth$ due to the limitations of our model which focuses on the interaction of gas and silicates.  Creating smaller planets would require planet core masses less than 4~$R_\Earth$, the smallest core mass in our simulations.  Accurately modeling the occurrence rate of smaller planets would require a model for the distribution of core masses and compositions.  Similarly, the predictions for radii larger than $\sim$5~$R_\Earth$ are not realistic, since our model does not include runaway accretion of gas once the atmosphere mass dominates the core mass. Runaway accretion will further depopulate the 3-6~$R_\Earth$ region of the plot, so runaway will result in a further decrease in the rate of these planets. 

\begin{figure*}[t]
(a)  \includegraphics[width=1.025\columnwidth,clip=true,trim={10 0 55 40}]{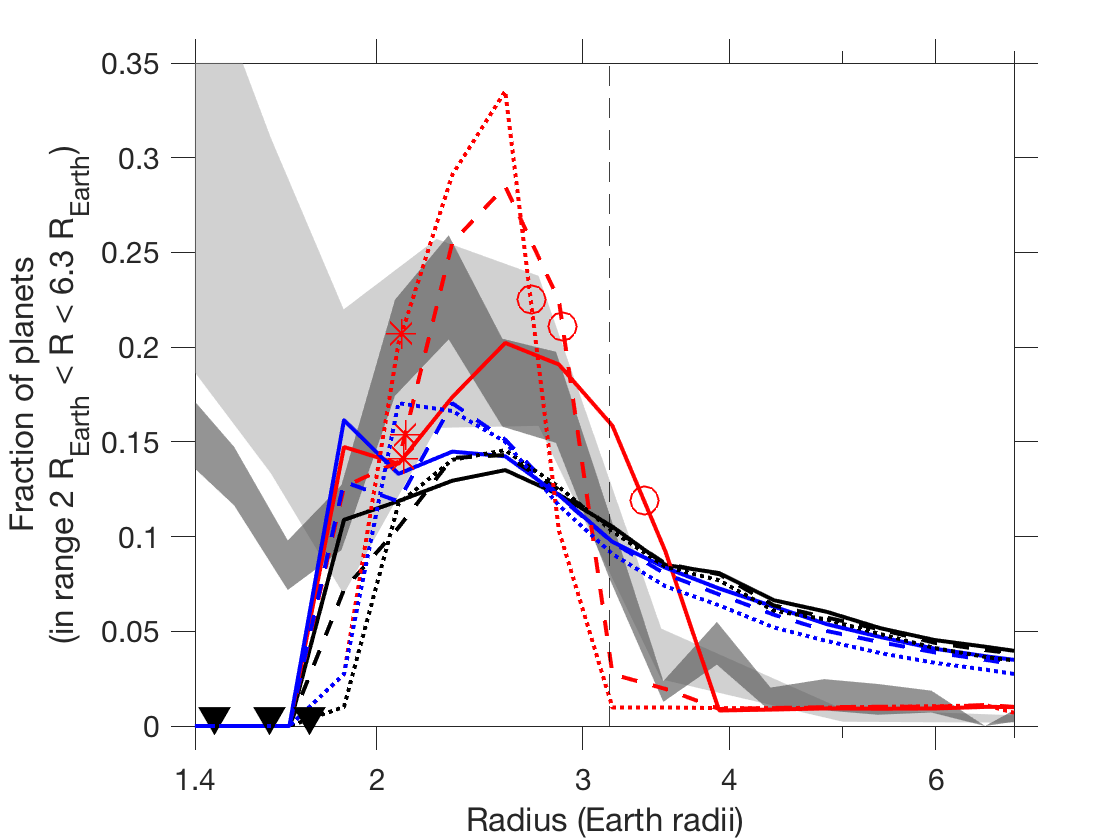}
(b) \includegraphics[width=1.025\columnwidth,clip=true,trim={10 0 55 40}]{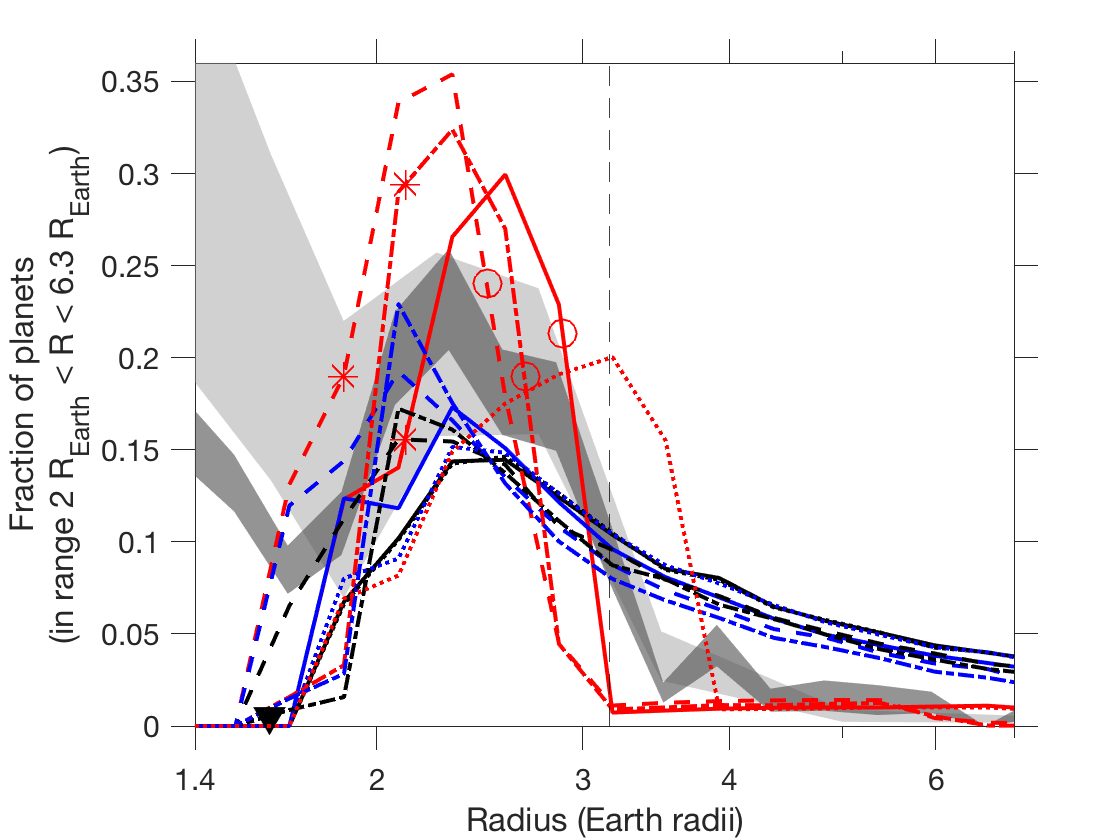} %sic (saved figure under wrong panel name - _4c_v3 is really panel b)
(c) \includegraphics[width=1.025\columnwidth,clip=true,trim={10 0 55 40}]{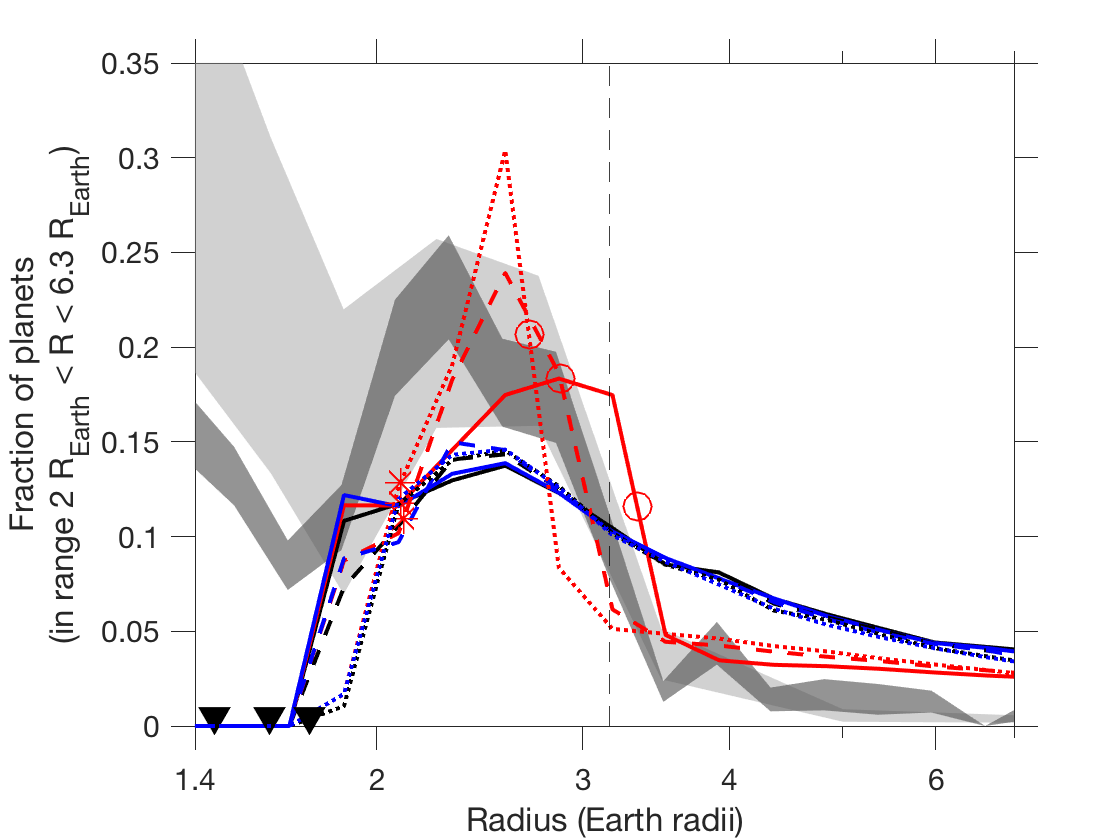}
(d) \includegraphics[width=1.025\columnwidth,clip=true,trim={10 0 55 40}]{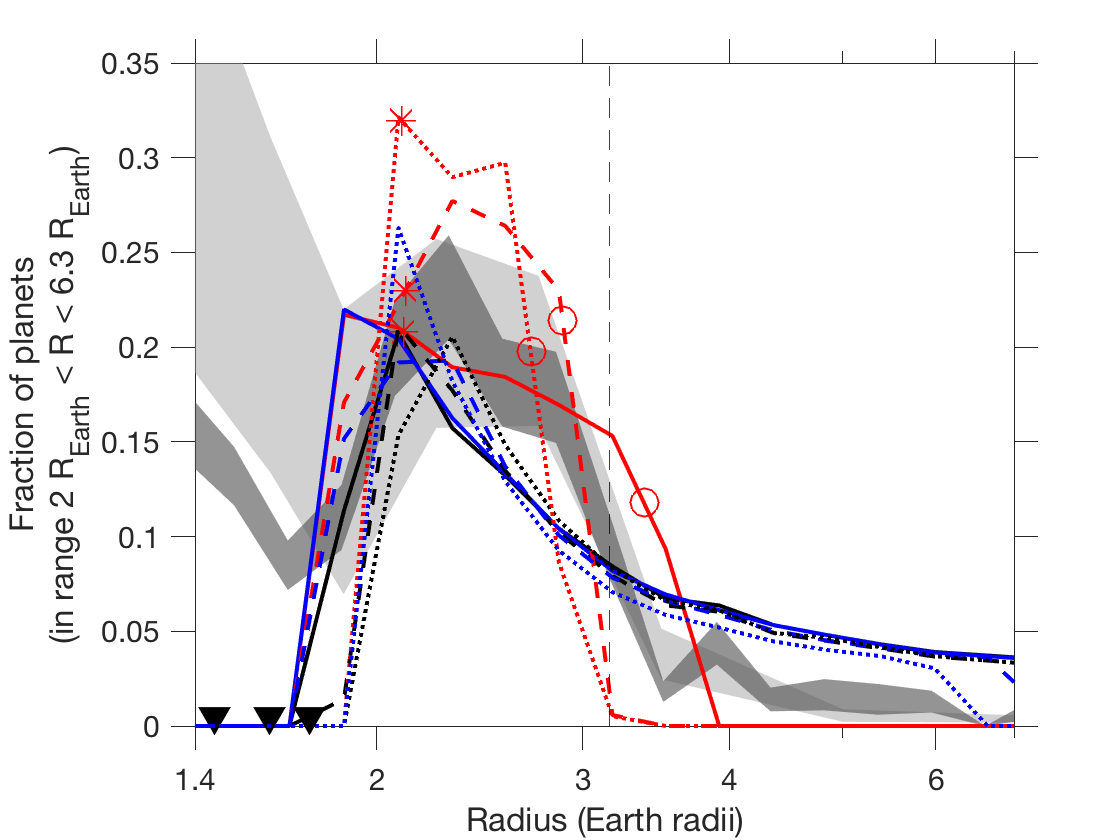}
\caption{ Histograms of planet abundance showing sensitivity to parameters. In each panel, the colored bands are the true planet histogram ($\pm$1$\sigma$ error) according to  \citet{Hsu2019} (light gray) and according to \citet{FultonPetigura2018} (dark gray). The black lines show the impermeable-planet cases. The blue lines show the linear (Henry's Law) dissolution cases. The red lines show the fugacity-crisis cases. For the red lines, the asterisks show atmosphere-base pressure~=~1~GPa and the open circles show $P_{atm}$~=~8~GPa. Model output bins are the same as in \citet{Fulton2017}, and the triangles correspond to the bare-core radii for the specified masses. (a) Sensitivity to core mass. Solid, dashed, and dotted lines are for 4~$M_\earth$, 6~$M_\earth$, and 8~$M_\earth$ respectively. This reference case shows model output for 1 Gyr, solar metallicity, $T_{mai}$~=~3000~K, insolation 1000 $L_{\Earth}$. (b)~Solid lines show the 6~$M_\Earth$ case from panel~a. The dashed lines vary $L$, which is set to 10 $L_\Earth$. The atmosphere is colder, but the magma is held at 3000~K. The dash-dot lines vary planet age, which is set to 10~Gyr. The dotted lines decrease the H$_2$ solubility by a factor of 10. (c)~As panel~a, but for $T_{mai}$~=~2500~K. (d)~As panel~a, but a log-uniform prior between bounds of 0.1~wt\%  and 30~wt\% H$_2$ added.}
\label{fig:sensitivity}
\end{figure*}

 Fig.~\ref{fig:sensitivity} shows the results of several variations on our reference model that serve as tests of the sensitivity of the fugacity crisis to model parameters. Fig.~\ref{fig:sensitivity}a shows that raising core mass from 4~$M_\Earth$ to 8~$M_\Earth$ shifts the cliff location by $\sim$0.4~Earth~radii. For smaller core masses the weight per unit mass of the atmosphere is less, so more H$_2$ mass can be added before reaching the limiting $P_{atm}$.

 Fig.~\ref{fig:sensitivity}b shows that decreasing insolation from 1000 $L_{\Earth}$ to 10 $L_{\Earth}$ shrinks the planets by 0.25 $R_\Earth$ (dashed lines). The effect of  increasing planet age from 1 Gyr to 10 Gyr (dash-dot line) is similar. Reducing H$_2$ solubility by a factor of 10 moves the cliff to larger radius by $\sim$0.7 $R_\earth$ (dashed line).

 Fig~\ref{fig:sensitivity}c shows results for $T_{mai}$~=~2500~K. The melt mass is greatly reduced (and insensitive to planet mass; 0.5-0.7~$M_\earth$ of melt for $M_{core}$~=~1-10~$M_\Earth$). This effect overpowers the greater solubility of H$_2$ in magma at lower $T$ (Fig.~\ref{fig:fugacitycrisis}a). Because there is less melt into which H$_2$ can dissolve, the amplitude of the cliff is reduced. However the results for incomplete melting are sensitive to the value of the maximum H$_2$ content of magma, which is poorly constrained. 
 
 Fig~\ref{fig:sensitivity}d shows the results for a log-uniform distribution of gas supply between bounds of 0.1~wt\% and 30~wt\% of core mass. The basic pattern is independent of choice of prior: the cliff gets steeper for non-ideal fugacity, and is especially steep for larger (8 $M_\Earth$) core masses. 
 
As shown in Fig.~\ref{fig:sensitivity}, the precise location and amplitude of the fugacity cliff depends on model parameters, including the distribution of core masses, atmosphere mass fractions, insolations, and $T_{mai}$. Nevertheless, these results show that a fugacity crisis is robust and can explain both the amplitude and the position of the radius cliff.

\section{Discussion.}

\subsection{Approximations and Limitations}
Our model provides a simple, equilibrium explanation for the radius cliff. However, this simplified model has limitations.

The most important limitation is the lack of H$_2$ solubility-in-magma data in the $\sim$4000~K and $10^9-10^{10}$~Pa regime of the sub-Neptune magma-atmosphere interface. This lack is understandable, because under such conditions magma is literally uncontainable (in that all material containers will melt). Nevertheless, more laboratory and/or numerical experiments are motivated. Meanwhile, we extrapolate from lower-temperature, lower-pressure data \citep{Hirschmann2012}. 

Another approximation is that we do not explicitly model partial molar volume of the dissolved gas. Including this effect would decrease solubilities and increase core volume, boosting planet radii. The addition of dissolved gas to the magma will also increase gravity throughout the atmosphere, which (because solubility depends on $P_{atm}$) shrinks radii. Determining which of these effects dominates would require a more sophisticated interior model.

Alternative choices for silicate composition could give a~solidus and liquidus hotter by up to 1000~K, curtailing melting \citep{Andrault2017}. On the other hand, real sub-Neptune $T_{mai}$ could be $>$4000~K, according to thermal models (e.g.~\citealt{HoweBurrows2015,Bodenheimer2018}). Moreover,  volatile addition favors melting; an effect we omit. Therefore, it is not clear whether or not our simple procedure overstates or understates magma mass. If the fugacity crisis hypothesis passes the tests we propose (\S4.2), then the need for some melt in order to sequester H could provide a~joint constraint on the temperature and silicate composition of cores.

As more atmosphere dissolves in the melt and vice versa, distinctions between melt and atmosphere must vanish. For example, reactions exemplified by 4H$_2$ + SiO$_{\mathrm{2(melt)}}$~=~SiH$_{4(gas)}$ + 2H$_2$O$_{\mathrm{(gas)}}$ can lead to partial dissolution of the cores in the atmosphere. This is potentially testable by observations of SiH$_4$ and other hydride/hydroxide gases of the rock forming elements. As the conditions for full magma-atmosphere miscibility emerge during planet formation, a fuzzy zone will develop at the magma-atmosphere interface. This zone is buoyant relative to the volatile-poor earlier-formed core. It is not known if convective transport through fuzzy zones is totally shut down, or merely reduced \citep{Garaud2018}. In either case, fuzzy-zone development at the atmosphere-core interface would restrict further dissolution of the atmosphere into the magma.

Our model ignores H$_2$O, so it does not apply to Neptune and Uranus, which are probably (although not certainly) H$_2$O-rich \citep{Helled2019}.

Although we assume the total amount of H$_2$ supplied by the nebula to the core is commonly in the range 0.1-100\% of core mass, our model says nothing about why this should be. Thus, our model complements studies of gas supply from the nebula to the core (e.g., \citealt{LeeChiang2016}).

\begin{figure}
 \includegraphics[width=1.02\columnwidth]{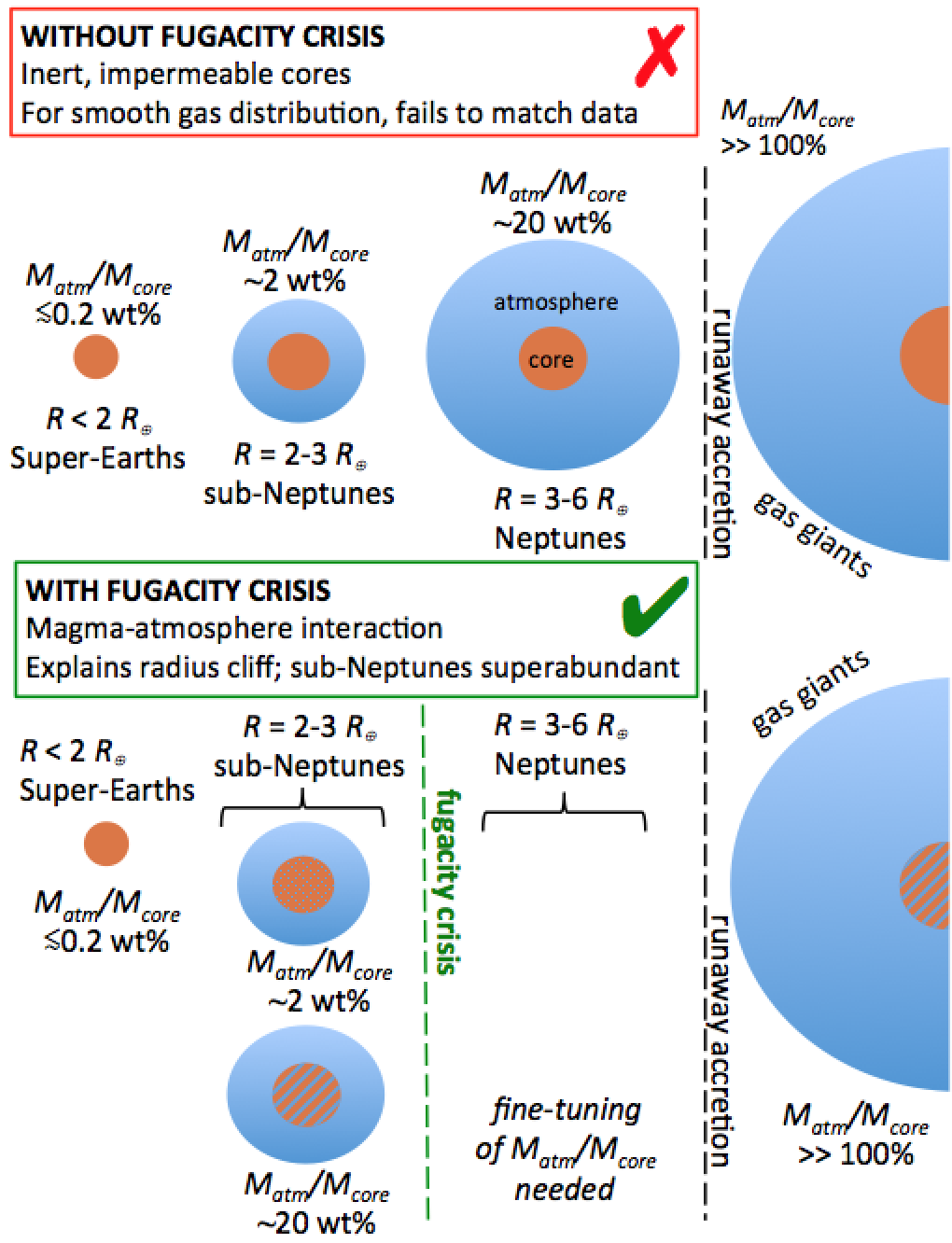}
\caption{Graphical abstract of this paper.}
\label{fig:cartoonsummary}
\end{figure}

\subsection{Alternatives and Tests}

If the magma ocean and the atmosphere equilibrate, then escape-to-space models understate the amount of gas that must be removed to affect planet radii. This is because gas loss will be compensated by exsolution -- a negative feedback (Fig.~\ref{fig:fugacitycrisis}b). This increases the energy demand on escape-to-space models. Moreover, if H$_2$ dissolves into magma then the planet's radius will be smaller during the crucial first 100~Myr, when the XUV flux is greatest.  In effect, the atmosphere hunkers down, reducing the number of hits from the XUV-photon fusillade. Moreover, the dissolved-in-magma H$_2$ will not be directly ejected by giant-impact shocks. These considerations suggest that \emph{if} the magma ocean and the atmosphere equilibrate, then the fugacity crisis is the only explanation for the cliff. 

The hypothesis of magma-atmosphere equilibration makes the following testable predictions.

\begin{enumerate}[leftmargin=*]
\itemsep0em

\item{\emph{Cliff Steepness.}} The fugacity crisis hypothesis is motivated by cliff steepness (Fig.~1). If more data and analysis makes this cliff less steep, that would not disprove the fugacity crisis, but it would dilute the attractiveness of this physics playing a dominant role in shaping final planet radii.

\item{\emph{Insensitivity To Formation Environment, Host Star Mass, e.tc.}}. As an equilibrium explanation, the fugacity crisis applies regardless of disk lifetime, host star mass, etc. Therefore, this model would be disfavored by a strong dependence of cliff location on star mass (for example).

\item{\emph{Atmospheric Chemistry}}. The fugacity crisis model requires a large amount of magma to interact with the atmosphere. Due to differential solubility (and likely partitioning of C into the Fe-metal-phase; \citealt{DasguptaGrawal2019}), this will drive elemental ratios (e.g. C/O) away from the solar value. This can now be tested (e.g. \citealt{Benneke2019}).

\item{\emph{Mass Dependence of Cliff Position.}} Our model predicts that more-massive planets (as a set) should have a cliff position that is at smaller radius than less-massive planets (as a set). This assumes that $f_{atm}$ is independent of core mass. This prediction is in tension with the radius analysis of \citet{Wu2019}. This motivates precision radial velocity surveys of a large number of planetary systems, ideally with multiple transiting planets.

\item{\emph{Gentler Slope For Longer Periods and Older Stars}}. The ensemble of sub-Neptunes with magma-atmosphere interfaces that are cold enough to crystallize (longer periods and older stars) should have a gentler cliff slope and thus a greater proportion of $\gtrsim$4~$R_\Earth$ planets. This is because core crystallization reduces the extent to which the magma can store hydrogen. This motivates future transit surveys that significantly increase the number of stars surveyed for transiting planets with equilibrium temperatures less than 400~K.

\end{enumerate}

\subsection{What Do Active Cores Imply?}
 
The growth process and the birth location for sub-Neptunes are unknown (e.g., \citealt{Rogers2011,ChiangLaughlin2013, ChatterjeeTan2014,Levison2015,Ormel2017,Mordasini2018,Brouwers2018}).

Our model assumes chemically (and thus thermally) active cores, with magma-atmosphere equilibration. Equilibration will happen if the silicates are delivered as planetesimals or as pebbles after the atmosphere has formed \citep{Bodenheimer2018}. If the silicates and gas are accreted on similar timescales, then interaction would occur at progressively higher pressures during planet formation. Our model is an equilibrium explanation which assumes that during or after planet growth this history of planet assembly is stirred away. Stirring need not be complete in order for the fugacity crisis to explain the radius cliff, because a little bit of magma can make a big difference (Fig.~\ref{fig:sensitivity}c).

Pursuit of these tests and implications will be aided by future extended missions for TESS \citep{Huang2018}; PLATO; ARIEL; and more radial-velocity data for sub-Neptunes.

\section{Conclusion.}

The major feature in the exoplanet radius distribution is the rapid decrease in the occurrence rate of planets as size increases from 3 to 3.5~$R_\earth$. This can be understood as a consequence of the nonideal increase in H$_2$ fugacity above 1~GPa. As the base-of-atmosphere pressure approaches 10~GPa ($\sim$3$R_\Earth$), more and more of the added H$_2$ goes into the magma and so the radius does not increase much. 

It follows that H$_2$ supply from the nebula can have a broad mass distribution and still match the observed radius histogram. A~world with $<$1~wt\% H$_2$ can lose its atmosphere and become a Super-Earth; a~world with a few~wt\% H$_2$ becomes a sub-Neptune; a world with $\sim$20~wt\%~H$_2$ \emph{also} becomes a sub-Neptune because of the fugacity crisis described above; and a~world with a ratio of H$_2$ mass to core mass of \emph{O}(100\%) undergoes gravitational runaway and becomes a gas giant (Fig.~\ref{fig:cartoonsummary}). The main strength of the fugacity crisis hypothesis is that it is an equilibrium explanation; it is less dependent on transients from formation-era processes, which are hard to constrain and thus test. The main weakness of our explanation is that it depends on a limited number of laboratory measurements of H$_2$ solubility. Better material properties data, including lab and numerical experiments relevant to solubilities under sub-Neptune conditions, are needed to build better models of sub-Neptune evolution (e.g., \citealt{Hirschmann2012,SoubiranMilitzer2015}).

The fugacity crisis defines the radius cliff and so explains why sub-Neptunes are so common while Neptune-sized planets are rare. Although our simple model suggests a solution to one of the puzzles posed by sub-Neptunes, overall, it is striking that the most common type of transiting planet remains so poorly understood.
\vspace{-0.2in}

\acknowledgments \noindent \emph{Acknowledgements.} 
We thank M.M.~Hirschmann, L.A.~Rogers, D.~Hsu, and B. Fulton. Grants: NASA (NNX17AC02G, NNX16AB44G), NSF (AST-1517541). E.B.F.~ acknowledges support from the Center for Exoplanets and Habitable Worlds, which is supported by The Pennsylvania State University, the Eberly College of Science, and the Pennsylvania Space Grant Consortium. 
\vspace{0.05in}

\noindent \emph{Code availability.} Everything used to make this paper can be obtained for unrestricted further use by emailing the lead author.

\renewcommand\thefigure{A\arabic{figure}}    
\setcounter{figure}{0}

\clearpage

\end{document}